\documentclass[letterpaper]{jpconf}
\usepackage{graphicx}
\begin{document}
\title{Lepton universality test with $K_{l2}$ decays at NA62 experiment}

\author{Spasimir Balev}

\address{Scuola Normale Superiore di Pisa and INFN, Sezione di Pisa}
\ead{Spasimir.Balev@cern.ch}

\begin{abstract}
The experiment NA62 at CERN collected a large sample of $K^\pm$ leptonic decays in order to perform precise test of lepton universality, by measuring the helicity suppressed ratio $R_K=\Gamma(K^\pm\rightarrow e^\pm\nu)/\Gamma(K^\pm\rightarrow \mu^\pm\nu)$. The preliminary result of the analysis based on $51,089$ $K^+\rightarrow e^+\nu$ candidates is $R_K=(2.500\pm 0.016)\times10^{-5}$, consistent with the predictions of the Standard Model.
\end{abstract}

\section{Introduction}

The ratio of kaon leptonic decay rates $R_K=\Gamma(K^\pm_{e2})/\Gamma(K^\pm_{\mu2})$ is known in the Standard Model (SM) with excellent precision due to cancellation of the hadronic effects: $R_K^{SM}=(m_e/m_\mu)^2\left(\frac{m^2_K-m^2_e}{m^2_K-m^2_\mu}\right)^2(1+\delta R_{QED})=(2.477\pm0.001)\times10^{-5}$~\cite{cir}, where $\delta_{QED}=(-3.78\pm0.04)\%$ is a correction due to the inner bremsstrahlung (IB) $K_{l2\gamma}$ process which is included by definition into $R_K$~\footnote{Unlike the structure dependent (SD) $K_{l2\gamma}$.}. Being helicity suppressed due to $V-A$ structure of the charged weak current, $R_K$ is sensitive to non-SM effects. In particular in MSSM it is possible non-vanishing $e-\tau$ mixing, mediated via $H^+$, which can lead to few percent enhancement of $R_K$~\cite{mas}.

The present world average of $R_K=(2.467\pm0.024)\times10^{-5}$ is composed of three measurements from 1970s~\cite{pdg} and the recent KLOE final result~\cite{amb}, leading to a precision of $\sim1\%$. The NA62 experiment collected data during 2007 and 2008 aiming to reach accuracy of $\sim0.4\%$. The preliminary result on partial data set is presented here.

\section{Experimental setup}

The NA62 experiment utilized the NA48/2 beam line~\cite{bline} and detector setup~\cite{nim} with optimizations for $K_{e2}$ data collection. The beam line of NA48/2 experiment is designed to deliver simultaneously $K^+$ and $K^-$, produced on a beryllium target from SPS primary protons. The beams of (74$\pm$2) GeV/$c$ momentum are selected by a system of magnetic elements. After cleaning, shaping and focusing, the beams enter 114 m long vacuum decay volume. The momenta of the charged decay products are measured by magnetic spectrometer consisting of four drift chambers (DCHs) and a dipole magnet. The resolution of the spectrometer is $\sigma(p)/p=1.0\%\oplus0.044\%p$ ($p$ in GeV/$c$). A scintillator hodoscope (HOD), located after the spectrometer, sends fast trigger signals from charged particles and measures their time with a resolution of 150 ps. The electromagnetic energy of particles is measured by a liquid krypton calorimeter (LKr), a quasi-homogeneous ionization chamber with an active volume of 10 m$^3$, 27 $X_0$ deep and segmented transversally into 13,248 cells ($2\times2$ cm$^2$ each). The energy resolution is
$\sigma(E)/E=0.032/\sqrt{E}\oplus0.09/E\oplus0.0042$ and the spatial resolution in the transverse coordinates $x$ and $y$ for a single electromagnetic shower
is $\sigma_x=\sigma_y=0.42/\sqrt{E}\oplus0.06$ cm ($E$ in GeV). A beam
pipe traversing the centers of the detectors allows undecayed beam particles and muons from decays of beam pions to continue their path in vacuum.

The $K_{e2}$ decays are selected by trigger requiring coincidence of hits in the HOD planes ($Q_1$ signal) together with sufficient energy deposit ($>10$ GeV) in LKr. The $K_{\mu2}$ events are selected by $Q_1$ signal, downscaled by a factor 150. Both triggers also use loose requirement on DCH multiplicity.

\section{Analysis strategy}
Due to the topological similarity of $K_{e2}$ and $K_{\mu2}$ decays a large part of the selection conditions are common for both decays. It is required a presence of single reconstructed charged track with momentum 15 GeV$/c < p < 65$ GeV$/c$ (the lower limit is due to the 10 GeV LKr energy deposit requirement in $K_{e2}$ trigger). The track extrapolated to DCH, LKr and HOD should be within their geometrical acceptances. The CDA between the charged track and the nominal kaon beam axis should be less than 1.5 cm. The event is rejected if a cluster in the LKr with energy larger than 2 GeV and not associated with track is present, in order to suppress the background from other kaon decays.

\begin{figure}[t]
\begin{minipage}{18pc}
\includegraphics[width=18pc]{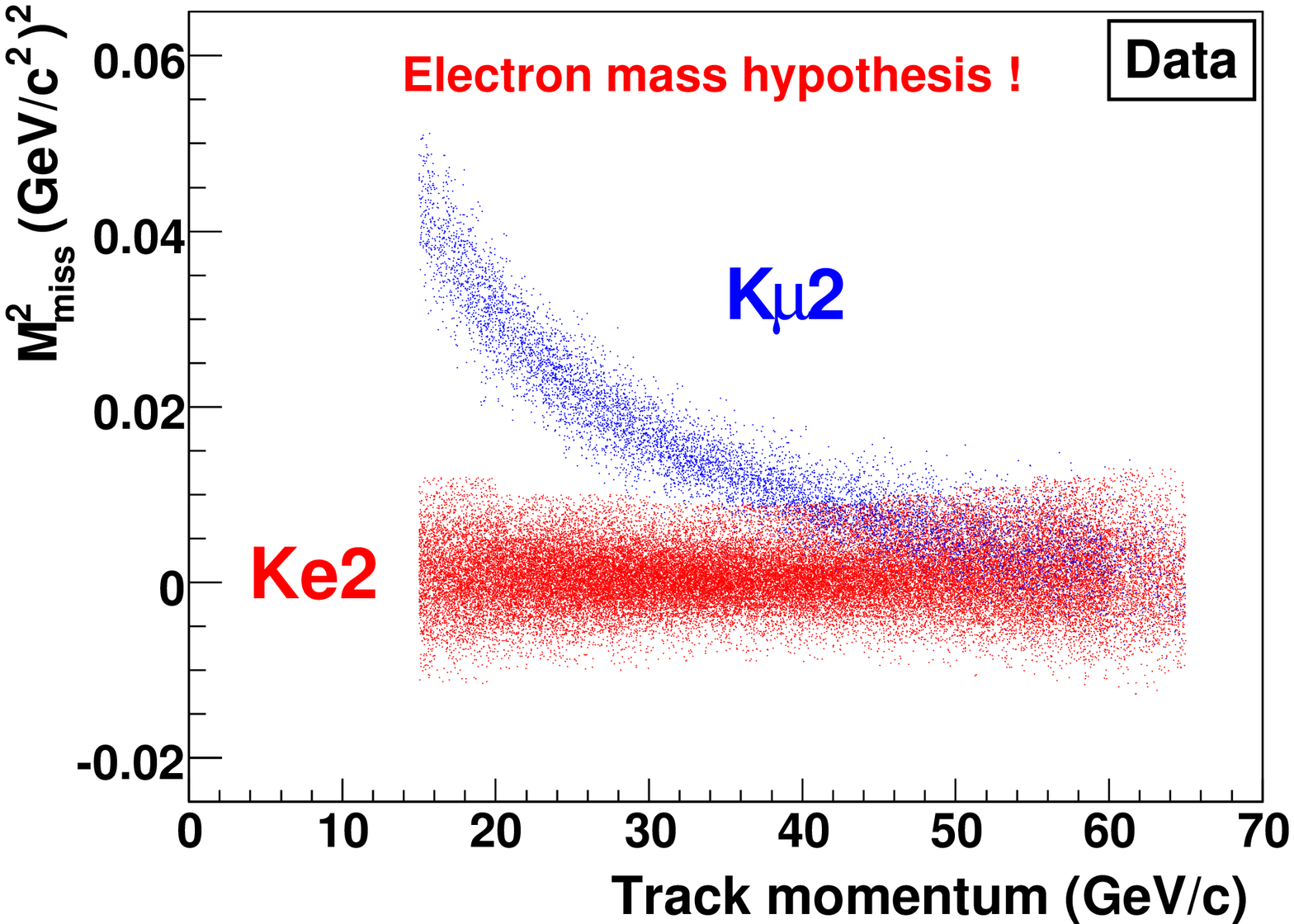}
\caption{\label{fig_sep}$M^2_{\rm miss}(e)$ for $K_{e2}$ and $K_{\mu2}$ decays.}
\end{minipage}\hspace{2pc}%
\begin{minipage}{18pc}
\includegraphics[width=18pc]{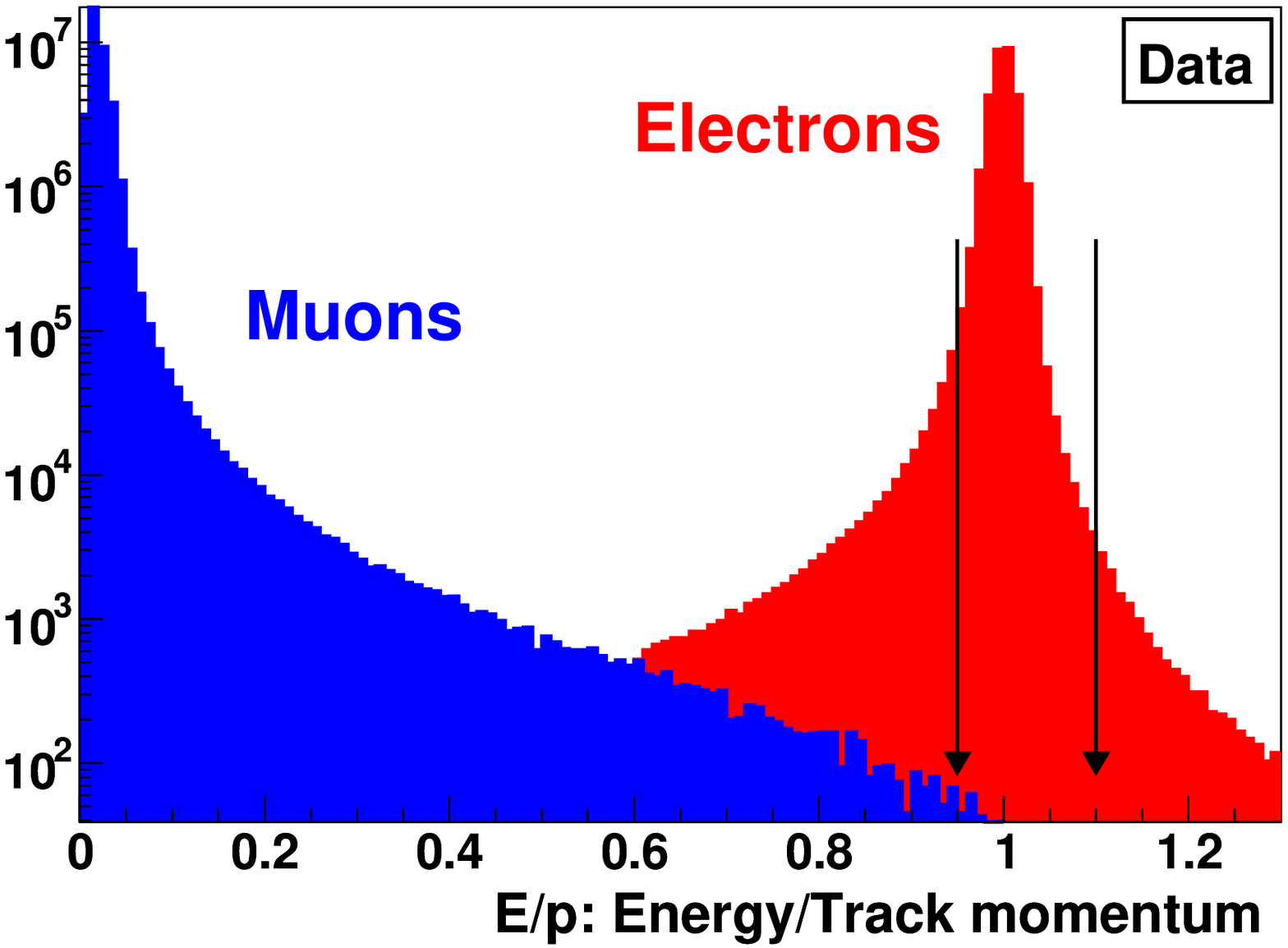}
\caption{\label{fig_eop}$E/p$ ratio for electrons and muons.}
\end{minipage} 
\end{figure}

A kinematical separation between $K_{e2}$ and $K_{\mu 2}$ for low track momenta is possible, based on the reconstructed missing mass, assuming the track to be an electron or a muon: $M_{\rm miss}^2(l)=(P_K-P_l)^2$, where $P_l$ ($l=e,\mu$) is the four-momentum of the lepton (see Fig.~\ref{fig_sep}). Since the kaon four-momentum $P_K$ is not measured directly in every event, its average is monitored in each SPS spill with fully reconstructed $K^\pm\rightarrow3\pi^\pm$ decays. A cut $|M^2_{\rm miss}(e)|<M_0^2$ is applied to select $K_{e2}$ candidates, and $|M^2_{\rm miss}(\mu)|<M_0^2$ for $K_{\mu2}$ ones, where $M_0^2$ varies from $0.009$ to $0.013$ (GeV/$c^2)^2$ for different track momenta, depending on $M_{\rm miss}$ resolution. Particle identification is based on the ratio $E/p$ of track energy deposit in the LKr to its momentum measured by the spectrometer. Particles with $0.95<E/p<1.1$ are identified as electrons, while particles with $E/p<0.85$ -- as muons (see Fig.~\ref{fig_eop}).

The analysis is based on counting the number of reconstructed $K_{e2}$ and $K_{\mu2}$ candidates with the selection described above. Since the decays are collected simultaneously, the result does not depend on kaon flux measurement and the systematic effects due to the detector efficiency cancel to first order. To take into account the momentum dependence of signal acceptance and background level, the measurement is performed independently in bins of reconstructed lepton momentum. The ratio $R_K$ in each bin is computed as
\begin{equation}
R_K=\frac{1}{D}.\frac{N(K_{e2})-N_B(K_{e2})}{N(K_{\mu2})-N_B(K_{\mu2})}.\frac{f_\mu\times A(K_{\mu2})\times\epsilon(K_{\mu2})}{f_e \times A(K_{e2})\times\epsilon(K_{e2})}.\frac{1}{f_{\rm LKr}},
\label{rk}
\end{equation}
where $N(K_{l2})$ are the numbers of selected $K_{l2}$ candidates ($l=e,\mu$), $N_B(K_{l2})$ are numbers of background events, $f_l$ are efficiencies of electron and muon identification criteria, $A(K_{l2})$ are geometrical acceptances, $\epsilon(K_{l2})$ are trigger efficiencies, $f_{\rm LKr}$ is the global efficiency of the LKr readout, and $D=150$ is the downscaling factor of the $K_{\mu2}$ trigger. In order to compute $A(K_{l2})$, a detailed Geant3-based Monte-Carlo simulation is employed. It includes full detector geometry and material description, stray magnetic fields, local inefficiencies, misalignment, detailed simulation of beam line, and time variations of the above throughout the running period.

\section{Backgrounds}

$N_B(K_{e2})$ in~(\ref{rk}) is dominated by $K_{\mu2}$ events with track misidentified as electron, mainly in case of high energetic bremsstrahlung after the magnetic spectrometer, when the photon takes more than 95\% of muon's energy. The probability for such process is measured directly by clean sample of muons passing $\sim10 X_0$ of lead (Pb) before to hit the LKr. A Geant4 simulation is used to evaluate the Pb correction to the probability for muon misidentification which occurs via two principal mechanisms: 1) muon energy loss in Pb by ionization, dominating at low momenta; 2) bremsstrahlung in the last radiation lengths of Pb increasing the probability for high track momenta. The background is evaluated to be $(6.28\pm0.17)\%$.

\begin{figure}[t]
\begin{minipage}{18pc}
\includegraphics[width=18pc]{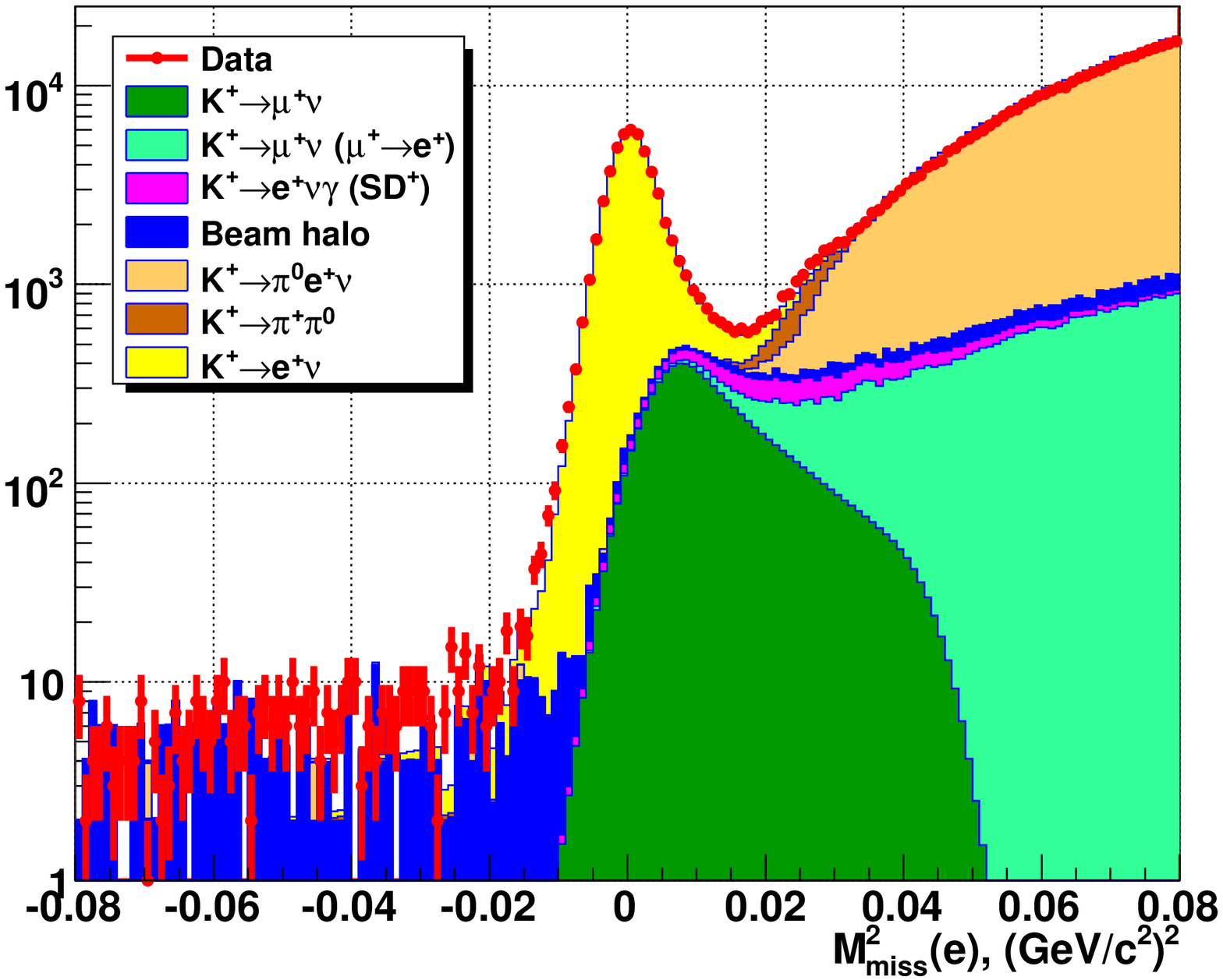}
\caption{\label{fig_mm2e_bg}$M^2_{\rm miss}(e)$ distributions for $K_{e2}$ candidates and for various backgrounds.}
\end{minipage}\hspace{2pc}%
\begin{minipage}{18pc}
\includegraphics[width=18pc]{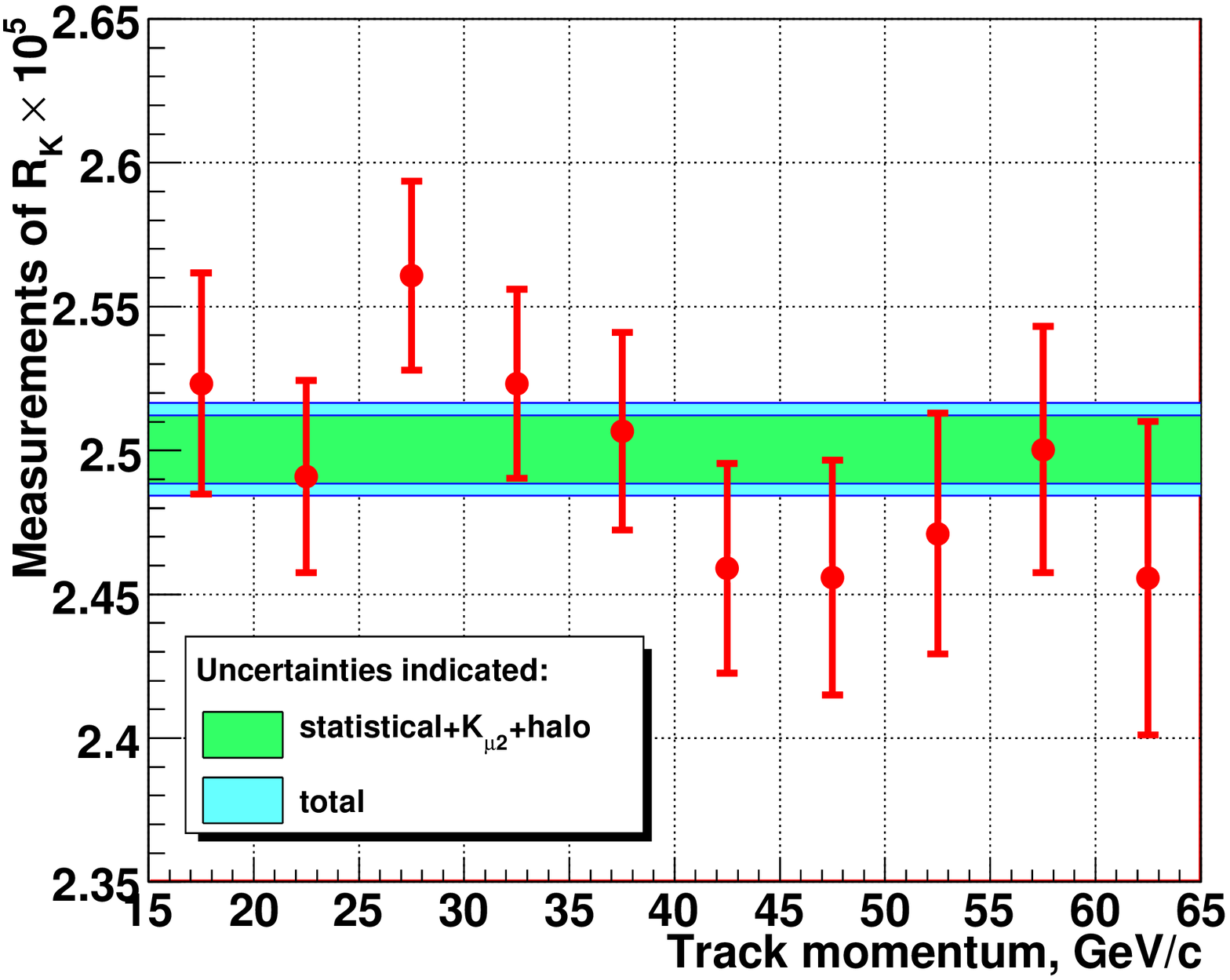}
\caption{\label{fig_rk}$R_K$ in track momentum bins.}
\end{minipage} 
\end{figure}

Since the incoming kaon is not tracked and the signature of $K_{l2}$ decays is a single reconstructed track, the background from beam halo should be considered. The performance of muon sweeping system results in lower background in $K_{e2}^+$ sample ($\sim1\%$) than in $K_{e2}^-$ sample ($\sim20\%$), therefore $\sim90\%$ of data were collected with the $K^+$ beam only, and small fractions were recorded with simultaneous beams and $K^-$ beam only. The halo background in $K^+_{e2}$ was measured to be $(1.45\pm0.04)\%$ directly by using data, collected when no $K^+$ beam is present. 

The other backgrounds considered are: $(0.23\pm0.01)\%$ from $K_{\mu 2}$ with subsequent $\mu\rightarrow e$ decay (suppressed according to Michel distribution, as muons from $K_{\mu 2}$ decays are fully polarized); $(1.02\pm0.15)\%$ from $K_{e2\gamma}$(SD) (the $\sim 15\%$ precision of the background evaluation is due to the experimental error on its branching~\cite{pdg}; the recent KLOE measurement~\cite{amb}, published after announcement of the NA62 preliminary result, is not taken into account); $0.03\%$ for both $K_{e3}$ and $K_{2\pi}$ decays.

The number of $K_{e2}$ candidates is 51,089 before background subtraction. The $M_{\rm miss}^2(e)$ distribution of data events and backgrounds are presented in Fig.~\ref{fig_mm2e_bg}.

\section{Systematic uncertainties and results}

The electron identification efficiency is measured directly as a function of track momentum and its impact point at LKr using electrons from $K_{e3}$ decays. The average $f_{e}$ is $(99.20\pm0.05)\%$ ($f_{\mu}$ is negligible). The geometric acceptance correction $A(K_{\mu2})/A(K_{e2})$ depends on the radiative $K_{e2\gamma}$(IB) decays, which are simulated in one-photon approximation~\cite{cir} without resummation of leading logarithms~\cite{gati}. The trigger efficiency correction $\epsilon(K_{e2})/\epsilon(K_{\mu2})\approx99.9\%$ accounts for the difference in the trigger conditions, namely the requirement of $E>10$ GeV energy deposited in LKr for $K_{e2}$ only. A conservative systematic uncertainty of $0.3\%$ is ascribed due to effects of trigger dead time. LKr global readout efficiency is measured to be $(99.80\pm0.01)\%$ using independent LKr readout. The various systematic uncertainties on $R_K$ result are summarized in Table~\ref{tab1}. 

\begin{center}
\begin{table}[h]
\centering
\caption{\label{tab1}Summary of the uncertainties in $R_K$ measurement in units of $10^{-5}$.} 
\begin{tabular}{@{}l*{15}{l}{l}{l}{l}{l}}
\br
Source&Error&Source&Error&Source&Error\\
\mr
Statistical&0.012&Beam halo&0.001&Geom. acceptance&0.002\\
$K_{\mu2}$ background&0.004&Electron ID&0.001&Trigger dead time&0.007\\
$K_{e2\gamma}$ background&0.004&IB simulation&0.007& & \\
\br
\end{tabular}
\end{table}
\end{center}

The independent measurements of $R_K$ in track momentum bins are presented in Fig.~\ref{fig_rk}. The preliminary NA62 result is $R_K=(2.500\pm0.012_{\rm stat}\pm0.011_{\rm syst})\times10^{-5}=(2.500\pm0.016)\times10^{-5}$, consistent with SM expectation. The analysis on the whole data set will allow to reach uncertainty of 0.4\%. The combined new world average is $(2.498\pm0.014)\times10^{-5}$.

\smallskip

\end{document}